\documentclass[fleqn,usenatbib,useAMS,linenumbers]{mnras}

   \makeatletter
\def\@fnsymbol#1{\ensuremath{\ifcase#1\or  \else\@ctrerr\fi}}
    \makeatother

\usepackage{subfigure}
\usepackage{fancyhdr}
\usepackage{booktabs}
\usepackage{graphicx}
\usepackage{threeparttable}
\usepackage{adjustbox}
\usepackage{csquotes}

\usepackage{graphicx}		
\usepackage{amsmath}		
\usepackage{amssymb}		
\usepackage{multicol}        	
\usepackage{bm}			
\usepackage{pdflscape}		

\usepackage[T1]{fontenc}
\usepackage{ae,aecompl}

\usepackage{newtxtext,newtxmath}

\title[]{Properties and Asteroseismological analysis of a new ZZ ceti discovered by TMTS}
 \author[Jincheng Guo et al.]{Jincheng Guo$^{1,\ast}$\thanks{$^{\ast}$\,E-mail: jcguo@bjp.org.cn (J,Guo), yanhuichen1987@126.com (Y,Chen), wang\_xf@mail.tsinghua.edu.cn (X.Wang)}, 
 Yanhui Chen$^{2,\ast}$,
 Xiaofeng Wang$^{3,1,\ast}$, 
 Jie Lin$^{3}$,
 Gaobo Xi$^{3}$,
 Jun Mo$^{3}$,\newauthor
 Alexei V. Filippenko$^{4}$,
 Thomas Brink$^{4}$,
 Xiao-Yu Ma$^{5}$,
 Weikai Zong$^{5}$, 
 Yong Yang$^{6,7}$,\newauthor 
 Jingkun Zhao$^{6,7}$,
 Xiangyun Zeng$^{8}$,
 Zhihao Chen$^{3}$,
 Ali Esamdin$^{9}$,
 Fangzhou Guo$^{3}$,\newauthor 
 Abdusamatjan Iskandar$^{9}$, 
 Xiaojun Jiang$^{6,7}$, 
 Wenxiong Li$^{10}$,
 Cheng Liu$^{1}$,
 Jianrong Shi$^{6,7}$,\newauthor 
 Xuan Song$^{1}$,
 Letian Wang$^{9}$,
 Danfeng Xiang$^{3}$,
 Shengyu Yan$^{3}$,
 Jicheng Zhang$^{5}$,\newauthor
 and Yonghui Yang$^{2,11}$
 \\
$^{1}$Department of Scientific Research, Beijing Planetarium, Xizhimenwai Street, Beijing 100044, China\\
$^{2}$Institute of Astrophysics, Chuxiong Normal University, Chuxiong 675000, China\\
$^{3}$Physics Department and Tsinghua Center for Astrophysics, Tsinghua University, Beijing 100084, China\\
$^{4}$Department of Astronomy, University of California, Berkeley, CA 94720-3411, USA\\
$^{5}$Department of Astronomy, Beijing Normal University, Beijing 100875, China\\ 
$^{6}$Key Laboratory of Optical Astronomy, National Astronomical Observatories, Chinese Academy of Sciences, Beijing 100012, China\\
$^{7}$University of Chinese Academy of Sciences, Beijing 100049, China\\
$^{8}$Center for Astronomy and Space Sciences, China Three Gorges University, Yichang, 443000, People's Republic of China \\
$^{9}$Xinjiang Astronomical Observatory, Chinese Academy of Sciences, Urumqi 830011, China \\
$^{10}$The School of Physics and Astronomy, Tel Aviv University, Tel Aviv 69978, Israel\\
$^{11}$Faculty of Science, Kunming University of Science and Technology, Kunming 650093, China\\
}

\begin{document}
 \date{}
 \pagerange{\pageref{firstpage}--\pageref{lastpage}} \pubyear{2022}
 \maketitle
 \label{firstpage}

\begin{abstract}
Tsinghua university-Ma Huateng Telescope for Survey (TMTS) aims to discover rapidly evolving transients by monitoring the northern sky. The TMTS catalog is cross-matched with the white dwarf (WD) catalog of Gaia EDR3, and light curves of more than a thousand WD candidates are obtained so far. Among them, the WD TMTS J23450729+5813146 (hereafter J2345) is one interesting common source. Based on the light curves from the TMTS and follow-up photometric observations, periods of 967.113\,s, 973.734\,s, 881.525\,s, 843.458\,s, 806.916\,s and 678.273\,s are identified. In addition, the TESS observations suggest a 3.39 h period but this can be attributed to the rotation of a comoving M dwarf located within $3^{''}$. The spectroscopic observation indicates that this WD is DA type with $T_{\rm eff}=11778\pm617$\,K, log\,$g=8.38\pm0.31$, mass=$0.84\pm0.20$\,M$_{\odot}$ and age=$0.704\pm0.377$\,Gyrs. Asteroseismological analysis reveals a global best-fit solution of $T_{\rm eff}=12110\pm10$\,K and mass=$0.760\pm0.005$ M$_{\odot}$, consistent with the spectral fitting results, and Oxygen and Carbon abundances in the core center are 0.73 and 0.27, respectively. The distance derived from the intrinsic luminosity given by asteroseismology is 93\,parsec, which is in agreement with the distance of 98\,parsec from Gaia DR3. Additionally, kinematic study shows that this WD is likely a thick disk star. The mass of its zero-age main-sequence mass is estimated to be 3.08\,M$_{\odot}$ and has a main-sequence plus cooling age of roughly 900\,Myrs.

\end{abstract}


 \section{Introduction}
In our Milky Way, up to 97\% of all stars will eventually evolve into white dwarfs (WDs) \citep{Heger2003}. Main-sequence stars with masses no more than 10-11\,M$_{\odot}$ \citep{Woosley2015} will end up as WDs, depending on their metallicity as well. WDs are objects of great importance that can provide vital information in variety of research fields \citep{Althaus2010}. Owing to a simple cooling mechanism, it is easier to derive relatively accurate ages of WDs, making them suitable for studies on the ages of Galactic populations \citep{Kilic2019,Guo2019}, which are important tools by studying their luminosity functions \citep{Munn2017,Lam2019} and mass functions \citep{Holberg2016,Hollands2018}. Accurate WD luminosity functions can be adopted to infer the structure and evolution of the Galactic disk and open and globular clusters \citep{Bedin2009,Campos2016,Garca2016,Kilic2017,Guo2018}. Furthermore, massive carbon-oxygen (CO) WDs are progenitors of type Ia supernovae (SNe Ia), which serve as standard candles for cosmological studies \citep{Maoz2014}. Ultra-short period double WDs are potential gravitational wave sources likely to be detected by space gravitational-wave facilities like LISA \citep{Burdge2019}.

With large spectroscopic surveys like Palomar-Green \citep{Green1986}, SDSS \citep{York2000} and LAMOST \citep{Zhao2012}, the studies of WDs have entered a new era by increasing the numbers of WDs greatly \citep{Kleinman2013,Kepler2015,Kepler2019,Zhao2013,Gentile2015,Guo2015,Guo2022,Kong2021}. Thanks to the Gaia mission \citep{Gaia2016}, the studies of WD now step into another new era. With the release of early Gaia DR3, the number of WDs reached $\sim$ 359\,000 \citep{Gentile2021}. This catalog is extremely helpful for studying light variability of WDs by cross-matching it with some photometric surveys.

Asteroseismology makes the studies of internal structure of pulsating stars become possible. This is also true for the studies of pulsating WDs (see a recent review by \cite{Corsico2019}). Since the first discovery of pulsating DA WDs (DAV, or ZZ ceti) HL Tau 76 \citep{Landolt1968}, there are as many as 494 DAVs identified via photometric observations \citep{Corsico2022}, including those discovered with TESS \citep{Romero2022}. Many of those DAVs have been analysed with asteroseismology from different groups using different models and asteroseismic approaches. Among them, two main methods are widely adopted. One uses static stellar models with parametrized chemical profiles. It allows a full exploration of the parameter space to find an optimal seismic model, leading to good matches to the observed periods. \citep{Bradley1998,Bradley2001,Castanheira2009,Bischoff2008,Bischoff2011,Bischoff2014,Giammichele2017a,Giammichele2017b}. Another approach is using full evolutionary models, which is generated by tracking the full evolution of the progenitor star, from the zero age main sequence to the WD stage \citep{Romero2012,Romero2013,Corsico2013,Geronimo2017,Geronimo2018}. It has been shown that detailed analysis on individual DAV is needed to understand the precise parameters of the white dwarf stars. In return, models can be improved based on observations. In general, the instability strip of DAVs is in the range of 10\,850\,K and 12\,270\,K \citep{Gianninas2011}, and the atmosphere of these cool WDs are almost dominated by pure hydrogen as a result of gravitational settling of other heavier elements. The hydrogen in the outer envelope recombines at around $T_{\rm eff}\sim 12\,000\,K$, which causes a huge increase in envelope opacity. Consequently, this restrains the flow of radiation and eventually results in  $g$-mode oscillations \citep{Winget1982}. Owing to the fact that these oscillation-induced pulsation modes are very sensitive to the stellar structure of DAVs, they can be used to study the internal structures of these stars. 

Over the past decade, the $Kepler$ \citep{Borucki2010} and its $K2$ mission \citep{Howell2014} have made high-precision asteroseismology of WDs possible \citep{Hermes2011,Greiss2016}. Later the $TESS$ mission \citep{Ricker2015} has enriched the sample, as it observes the whole sky \citep{Campante2016,Romero2022}. However, $Kepler$ and $K2$ mainly focus on specific areas; while $TESS$ will cover the whole sky area, but its size is too small for faint objects. Therefore, ground-based survey telescope with short cadence is still vital for discovery and asteroseismological study of faint pulsating WDs.

In this work, we report our discovery and results of asteroseismological analysis of a new DAV star TMTS J23450729+5813146 (dubbed as J2345) discovered by TMTS. This star, with coordinates of RA=23:45:07.29 and Dec=+58:13:14.6, was first reported in Gaia DR2 WD catalogue as a WD candidate with a probability being WD P$_{\rm WD}$=0.99 and G=16.88\,mag \citep{Gentile2019}. Through SED fitting to hydrogen-rich and helium-rich atmosphere models, the parameters of J2345 are estimated to be $T_{\rm eff}$=11\,280\,K, log\,$g$=8.04, mass=0.63 and $T_{\rm eff}$=11\,069\,K, log\,$g$=7.93, mass=0.54, respectively. Then it was studied in Gaia 100\,pc infrared-excess WD sample \citep{Rebassa2019}, as it seems to comove with a nearby faint (G=19.15\,mag) M star. In their paper, the effective temperature of J2345 and the companion are estimated to be 11\,750\,K and 2\,800\,K, respectively, via SED fitting. No public spectrum data is available for J2345 so far. 

This paper is arranged as follows: In Sect.\,2, observational details of J2345 are listed. In Sect.\,3, frequency solution and asteroseismological modelling of J2345 are presented. Period from TESS is analyzed in Sect.\,4, while discussions and summary are given in Sect.\,5 and 6, respectively.

\section{Observations}
\subsection{TMTS}

\begin{figure*}
    \centering
    \includegraphics[width=1.0\textwidth]{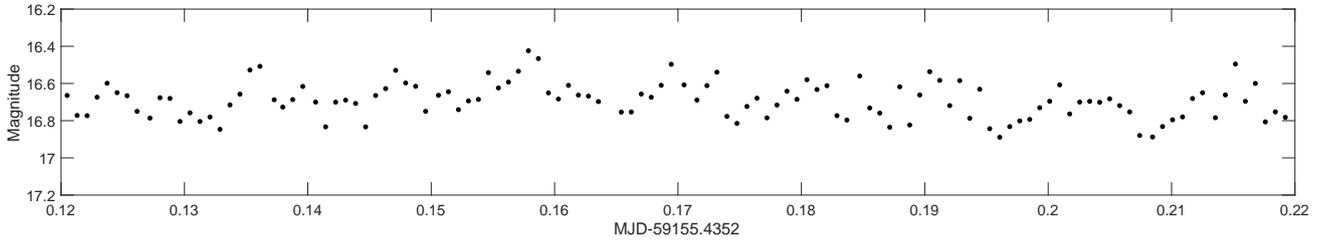}
    \caption{A part of the discovery TMTS light curve taken on December 2nd, 2020. The amplitude of variation is large for a DAV.}
    \label{tmts_lc}
\end{figure*}

TMTS is a photometric survey with four 40-cm optical telescopes, located at Xinglong station in China. It has a total field of view of $\sim$ 18\,deg$^{2}$ (4.5\,deg$^{2}$ each) and equipped with four 4k $\times$ 4k pixels CMOS cameras, which allow less than a second read-out time and high speed photometry. The survey operates in two observation modes. First, uninterrupted observing LAMOST areas for the whole night with a cadence of 1 min, carrying out collaborative tasks with the LAMOST. Second, supernova survey with a cadence of 1-2\,d. For the purpose of reaching out the full potential of observations, customized filter covering 330-900\,nm is used for the survey. The 3$\sigma$ detection limit of TMTS can reach 19.4\,mag for 1-min exposure. More description regarding the performance of TMTS can be found in \cite{Zhang2020}, \cite{Lin2022a} and \cite{Lin2022b}.

TMTS started the 1 min-cadence survey since 2020, producing uninterrupted light curves for more than 10 million objects so far. The WD catalog generated from Gaia EDR3 \citep{Gentile2021} is used to cross-match with the TMTS catalog within 3$^{\prime\prime}$, which yields more than a thousand WD common sources. Based on visual inspections of each source, WD J2345 was selected as a pulsating WD candidate for detailed study. WD J2345 was observed by TMTS on two nights (see details in Table.\,\ref{obslog}). A part of the discovery light curve is shown in Fig.\ref{tmts_lc}. The amplitude of variation is large for a ZZ ceti, roughly 0.2-0.3\,mag.

\begin{table}
\footnotesize
\begin{center}
\caption{Observation logs of J2345.}
\label{obslog}
\begin{tabular}{lllll}
\\ \hline
Telescope & instrument & obs\_date & exposure (s) \\   \hline
TMTS & CMOS & 2020-11-01 & 60    \\
(4$\times$40cm)     & (330-900\,nm) & 2020-11-02 & 60      \\
\\
SNOVA (40cm)& CCD & 2021-11-07 & 50      \\
      & (White light) & 2021-11-08 & 100   \\
      &           & 2021-11-10 & 100        \\
      &           & 2021-11-11 & 100       \\
      &           & 2021-11-14 & 100     \\
\\
Shane (3m) & Kast Double  & 2021-11-07 &   3600  \\
                     &    Spectrograph  & 2021-11-12 & 2400     \\
\\
TESS &  CCD    & 2019/Sector17  & 120/1800       \\   
     &   (600-1000\,nm)   & 2020/Sector24  & 120/1800       \\  \hline
\end{tabular}
\end{center}
\end{table}

\begin{figure}
    \centering
    \includegraphics[width=0.55\textwidth]{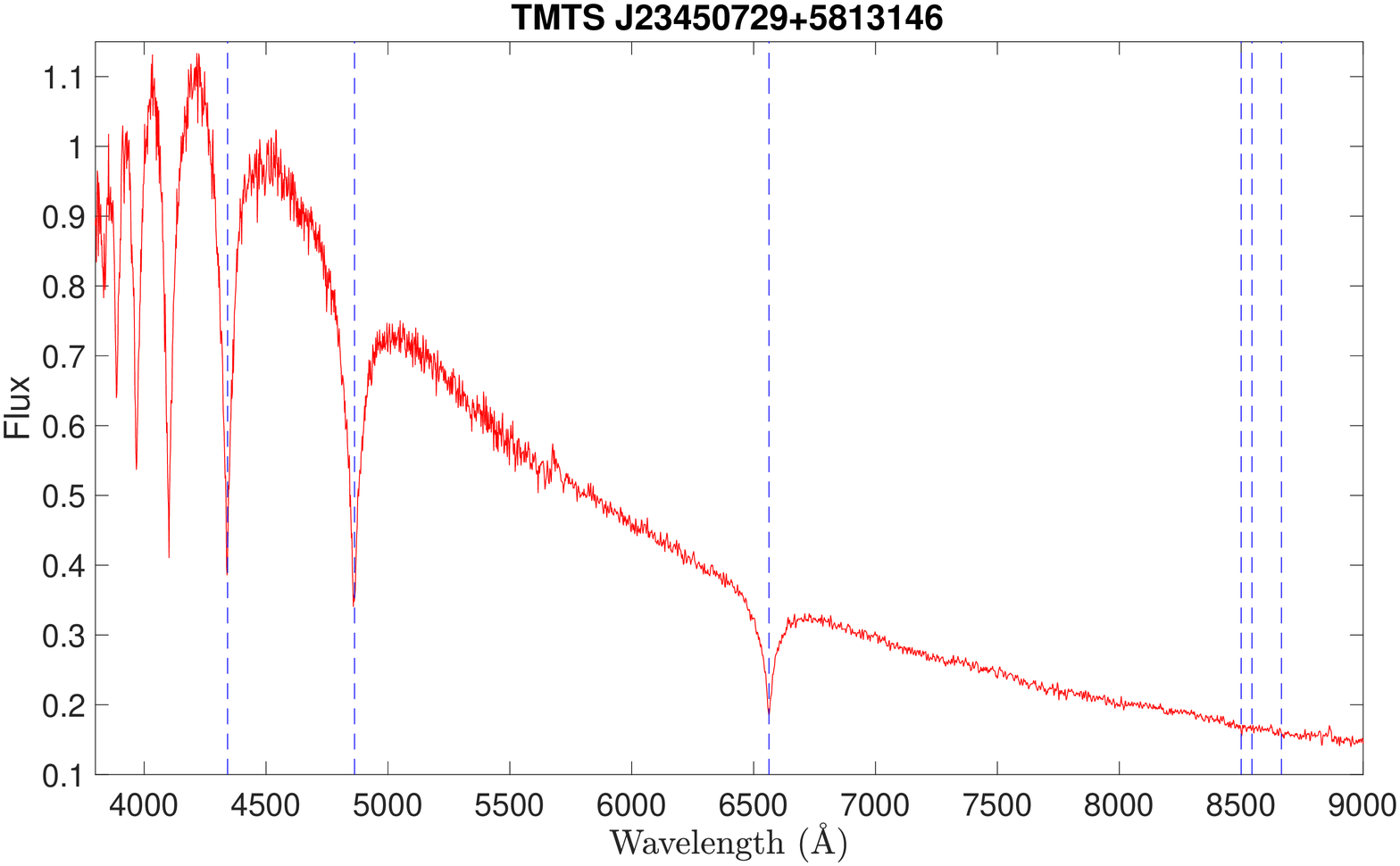}
    \includegraphics[width=0.4\textwidth]{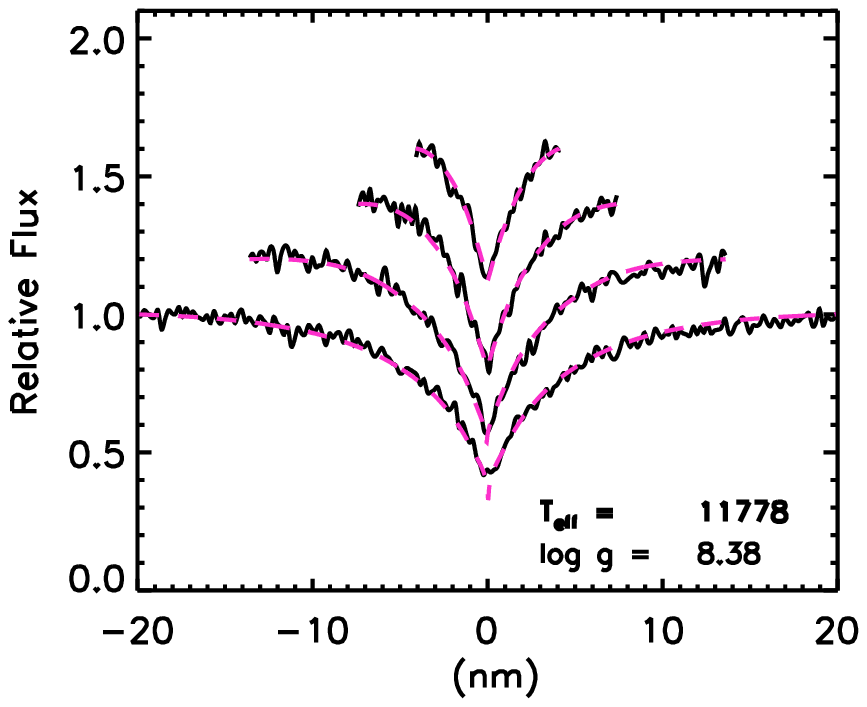}
    \caption{Spectrum taken by Lick 3\,m Shane telescope and its spectral fitting. From left to right, blue dashed lines in the upper panel mark the H$_{\gamma}$, H$_{\beta}$, H$_{\alpha}$, and Ca II triplet lines, respectively. From bottom to top, black lines in the lower panel are normalized H$_{\alpha}$, H$_{\beta}$, H$_{\gamma}$, and H$_{\epsilon}$ lines, respectively. Red dashed lines are the best model fitting after corrections for blueshift.}
    \label{spectrum}
\end{figure}

\subsection{Spectral observations}
Spectral follow-up observations have been carried out by 3\,m Shane telescope located at Lick observatory in 2021. Both spectra were taken by Kast double spectrograph with slit width of 2$^{''}$ and grism 2, covering 3600-5700\,\AA\,in blue and 5400-10760\,\AA\,in red, providing a spectral resolution of roughly 1.02\,\AA$\cdot pixel^{-1}$. The standard procedures of dark current subtraction, bias subtraction, cosmic ray removal, 1D spectral extraction are performed using IRAF. The signal to noise ratio (SNR) of the spectrum taken on 12th November is much better than the other due to better seeing, thus only the spectrum with higher SNR is used for spectral analysis. Owing to its higher SNR, blueshifted lines are noticeably seen in the spectrum. Spectral fitting with \texttt{WDTOOLS} \citep{Chandra2020,Chandra2020a} yields a radial velocity of -67.7$\pm$17.2\,km$\cdot s^{-1}$. After corrections for the line shift, the parameters of J2345 are derived as $T_{\rm eff}=11778\pm617$\,K, log\,$g=8.38\pm0.31$, mass=$0.84\pm0.20$\,M$_{\odot}$ and age=$0.704\pm0.377$\,Gyrs, based on the spectral model fitting method described in \cite{Guo2015,Guo2022}. The spectrum and best fitting are shown in Figure.\,\ref{spectrum}. The derived $T_{\rm eff}$ and mass are especially helpful for providing independent check of the results from asteroseismological analysis.

\subsection{Photometric follow-up}
Additional follow-up photometric observations have been taken by a 40\,cm telescope called SNOVA located at Nanshan station in China. A total of 5 nights observations have been taken for J2345. All these observations are observed in white light with exposure time of around 100\,s. 
The standard image processing, like bias correction, flat correction and source extraction are performed with IRAF, then differential photometry is applied to obtain the final light curves. In addition, TESS has also observed J2345 in 2 sectors in 2019 and 2020 (see Table.\,\ref{obslog}). Analysis of the light curves is presented in section 4.

\section{Analysis}

\begin{table*}
	\centering
	\caption{The frequency content detected from TMTS and SNOVA photometry and their mode identifications.}
	\label{tab:1_table}
    \small
    \begin{tabular}{cccccccccccc}
    	\hline
    	ID & Freq.           & $\sigma {\rm Freq.}$ & $\rm P$ & $\sigma$ P & Amplitude & $\sigma \rm{Amp}$ & SNR & Optimal P$_{\star}$ & $(l,k)_{\star}$ & Optimal P$_{\dagger}$ & $(l,k)_{\dagger}$ \\
	    & ($\mu$Hz)  &  ($\mu$Hz)    &  (s) &  (s) &  (ppt) &    (ppt) &     &   (s) &   & (s) &  \\ \hline
	    &            &               &     &   &     TMTS 2020  &   &         &       &\\
	     \\
        $f_{1}$&1026.974&0.587 &973.734&0.556&69.326&7.973&8.7& 973.276   & (1,19)    &   973.751 & (1,19) \\  \hline
            &  &    &     &   &    SNOVA 2021   &          &    &    &  &  &  \\  
              \\
		$f_{05}^*$   &  1034.005  &   0.262  &      967.113  &   0.245  &   13.128  &   3.980  &    3.3   &967.245  & (2,34) & 967.061 & (2,34) \\ 
        $f_{03}$   &  1134.398  &   0.218  &      881.525  &   0.169  &   15.106  &   3.806  &    4.0   & 881.448 & (1,17) & 881.556 &  (2,31)\\ 
        $f_{02}$   &  1185.596  &   0.205  &      843.458  &   0.146  &   15.944  &   3.769  &    4.2  & 843.661  & (1,16)  & 843.394 & (1,16) \\ 
        $f_{04}^*$   &  1239.286  &   0.245  &      806.916  &   0.159  &   13.409  &   3.789  &    3.5  &  806.992  & (1,15)  &  806.855 & (1,15)\\ 
        $f_{01}$   &  1474.333  &   0.061  &      678.273  &   0.028  &   45.447  &   3.217  &   14.1   & 678.358  & (1,12)    &  678.243  & (2,23)\\   \hline
	\end{tabular}	
	\begin{tablenotes}
	    \item $^{*}$ Frequencies below the 4.0\,$\sigma$ detection limits.
	    \item $_{\star}$ Adopted solution.
	    \item $_{\dagger}$ Additional solution found by freeing the value of $l$.
	\end{tablenotes}
\end{table*}

\begin{figure}
    \centering
    \includegraphics[width=0.5\textwidth]{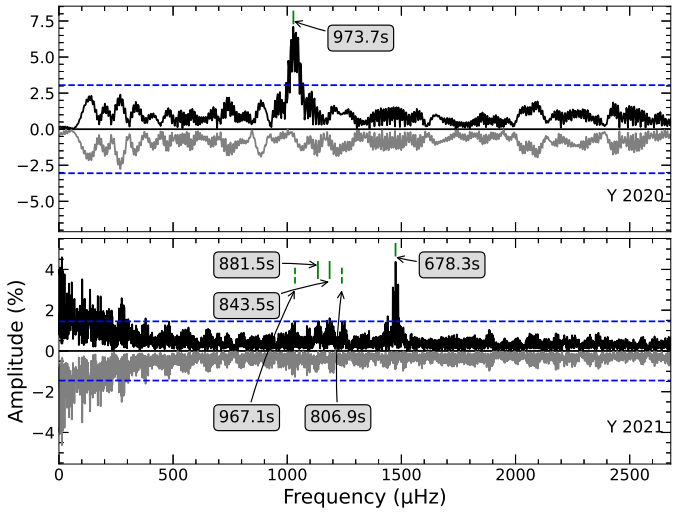}
    \caption{The Lomb-Scargle periodogram of TMTS and SNOVA observations taken in 2020 and 2021, respectively. The horizontal dashed lines (in blue) denote the 4.0\,$\sigma$ significance threshold. The upside-down grey curves represent the residuals of power spectra with significant frequencies being pre-whitened.}
    \label{fig:FT}
\end{figure}

\subsection{Frequency content}

The flexible software package \texttt{FELIX}\footnote{Frequency Extraction for Lightcurve exploitation, which was developed by St\'{e}phane~Charpinet. See details of the code in \citep{Charpinet2010}.} was used for Fourier decomposition of the light curve, which incorporates the standard prewhitening and nonlinear least-square fitting techniques for Fourier transformation \citep{Deeming1975,Scargle1982}. The threshold of acceptance for a significant frequency is set to be 4.0\,$\sigma$ of the local noise level \citep[see, e.g.,][]{1997A&A...328..544K}. Table~\ref{tab:1_table} lists the frequencies with their attributes extracted with \texttt{FELIX}: the ID ordered by their frequency (Column~2), amplitude (Column~6) and their uncertainties (Columns~3, 7), the corresponding period and their uncertainties (Columns~4, 5), the SNR (Column~8), the corresponding periods of optimal model, which will be discussed in Section~\ref{sec:fitting} (Column~9) and their $l,k$ (Column~10). We have barely detected one significant frequency, $f_1 \sim 1027 \mu$\,Hz in the year of 2020 from TMTS, while three significant frequencies and two suspected frequencies from SNOVA in the year of 2021.

Figure.\,\ref{fig:FT} shows the Lomb-Scargle periodogram of TMTS and SNOVA observations with one and five (including 2 frequencies below the detected limitation) frequencies detected. We clearly see that the significant frequency in 2020 with period of about 973~s disappeared in 2021. However, the five frequencies with period from 678~s to 967~s in 2021 were also not detected in 2020. This indicates that significant frequency and amplitude variations happened in J2345. Note that this phenomenon has been disclosed in pulsating white dwarf stars from both ground-based and space-borne photometry \citep[e.g.,][]{2011A&A...528A...5V,Zong2016}. A natural explanation could be that the nonlinear mode interactions between resonant modes \citep{Zong2016,1995A&A...296..405B}. We thus suggest that all of the above six signals are real pulsations and will be used for construction of seismic models. 

\subsection{Input physics and model calculations}
The latest version of the White Dwarf Evolution Code \cite[\texttt{WDEC} v16,][]{Bischoff2018a} is used in this work to evolve grids for DAV models. This code includes the oxygen (O) profile in the core, parameterized by six parameters (i.e. h1-3, w1-3). This feature provides more accurate fitting results than previous version without it. The equation of state and opacity tables from the Modules for Experiments in Stellar Astrophysics \cite[MESA, ][]{Paxton2011} are adopted in current version of \texttt{WDEC}. The standard mixing length theory \citep{Bohm1971} and the mixing length parameter of 0.6 \citep{Bergeron1995} are adopted in the analysis. The explored parameter spaces of evolved WD models are shown in Table.\,\ref{ParaSpace}. For the initial ranges and crude steps in this table, 7\,558\,272 WD models have been evolved. The model adopted here, differs from those derived by considering the complete evolution. Firstly, this model only considers the cooling phase of WD stars, while fully evolutionary models take into account the complete evolution of the progenitor stars and the WD cooling phases. Secondly, this model here is a large sample model grids including parameterized central core, while fully evolutionary models are typically not used for large sample model fitting. More details regarding the core parameters can be found in the Fig.\,1 of \cite{Bischoff2018a} and the user manual \citep{Bischoff2018b}. In the fitting process, fine steps are used to optimize the fitting models.

\begin{table*}
\begin{center}
\caption{The explored parameter spaces of WDs evolved by \texttt{WDEC}.}
\label{ParaSpace}
\normalsize
\begin{tabular}{lllllllllll}
\hline
Parameters                             &Initial ranges     &Crude steps    &Fine steps     &Optimal values$_{\star}$      & Optimal values$_{\dagger}$   \\
\hline
$M_{*}$/$M_{\odot}$                    &[0.500,0.850]      &0.010          &0.005          &0.760$\pm$0.005     & 0.690$\pm$0.005    \\
$T_{\rm eff}$(K)                       &[10600,12600]      &250            &10             &12110$\pm$10              &  12370$\pm$10    \\
-log($M_{\rm env}/M_{\rm *}$)          &[1.50,2.00,3.00]        & 0.50/1.00           &0.01           &2.00$\pm$0.02         &  3.00$\pm$0.01  \\
-log($M_{\rm He}/M_{\rm *}$)           & [2.00,5.00]        &1.00           &0.01           &5.01$\pm$0.04          &  4.99$\pm$0.01 \\
-log($M_{\rm H}/M_{\rm *}$)            &[4.00,10.00]       &1.00           &0.01           &9.98$\pm$0.02          &  7.00$\pm$0.01 \\
$X_{\rm He}$ in mixed C/He/H region    &[0.10,0.90]        &0.16           &0.01           &0.25$\pm$0.01    &  0.56$\pm$0.01     \\
\hline
$X_\mathrm{O}$ in the core             &                   &               &               &                       &   \\
\hline
h1                                     &[0.60,0.75]        &0.03           &0.01           &0.73$\pm$0.01    &  0.66$\pm$0.01       \\
h2                                     &[0.65,0.71]        &0.03           &0.01           &0.66$\pm$0.01     & 0.71$\pm$0.01      \\
h3                                     &0.85               &               &0.01           &0.81$\pm$0.03           & 0.85$\pm$0.01    \\
w1                                     &[0.32,0.38]        &0.03           &0.01           &0.35$\pm$0.01    &  0.37$\pm$0.01     \\
w2                                     &[0.42,0.48]        &0.03           &0.01           &0.45$\pm$0.01    &  0.46$\pm$0.01       \\
w3                                     &0.09               &               &0.01           &0.09$\pm$0.01          &  0.12$\pm$0.01      \\
\hline
\end{tabular}
\end{center}
\begin{tablenotes}
\item $M_{*}$ denotes the stellar mass, while $M_{env}$ is the envelop mass. $M_{H}$ is the mass of hydrogen atmosphere, and $M_{He}$ is the mass of He layer. $X_{He}$ is the He abundance in the mixed C/He/H region, and $X_{O}$ is the O abundance in the core. The parameter h1 refers to the O abundance in the core center, while w1 is the mass fraction of $X_{O}$=h1. Parameters h2 and h3 are O abundance of two knee points on the reducing O profile. The w2 and w3 refer to the masses of the gradient regions of O profile. 
	    \item $_{\star}$ Adopted solution.
	    \item $_{\dagger}$ Additional solution found by freeing the value of $l$.
\end{tablenotes}

\end{table*}

\begin{figure*}
    \centering
    \includegraphics[width=0.45\textwidth]{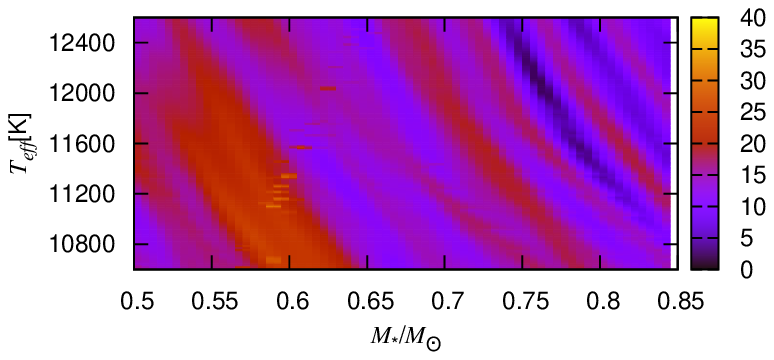}
    \includegraphics[width=0.45\textwidth]{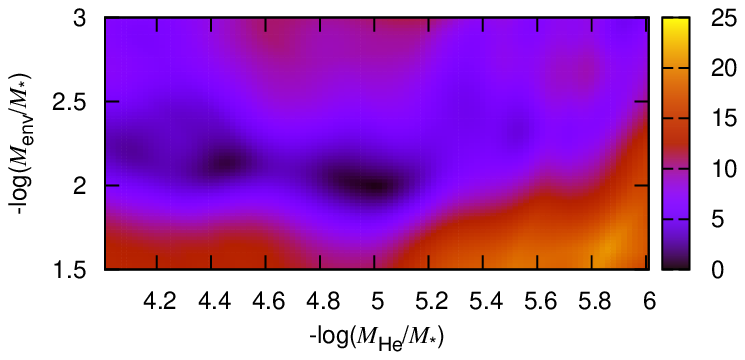}
    \caption{Left panel: a color graph of fitting error $\sigma_{RMS}$ as functions of stellar mass and $T_{\rm eff}$. A total of 14\,271 DAV stellar models are used in fit. Right panel: a color graph of fitting error $\sigma_{RMS}$ as functions of He-layer mass and envelope mass. A total of 7\,676 DAV stellar models are used in the fit. }
    \label{fig:teffmass_envhe}
\end{figure*}

\subsection{The model fittings}
\label{sec:fitting}

For the model fittings, a root-mean-square residual $\sigma_{RMS}$ is adopted in this work for the evaluation of fitting results.
\begin{equation}
\sigma_{RMS}=\sqrt{\frac{1}{n}\sum_{n}^{}(P_{obs}-P_{cal})^{2}}
\label{eq:1}
\end{equation}
In Eq.\ref{eq:1} above, $n$ is the number of fitted observed modes, i.e., 6 for J2345. $P_{obs}$ and $P_{cal}$ are the observed periods and model calculated periods, respectively. During model fittings, only 2 digits behind decimal points of observed periods are used. Owing to the fact that periods 973.73\,s and 967.11\,s are close and the amplitude of period 973.73\,s is large, we assume the period of 967.11\,s is $l$=2 mode, and the other 5 periods belong to the $l$=1 mode. Additional analysis has been checked as well, leaving free the value of $l$ for all periods. For the initial fitting, the combinations of parameter space ranges and crude steps are used to build the DAV models. This includes 7\,558\,272 models, which are adopted to fit the observed 6 modes. The crude steps are chosen by considering the computational requirement for models. It takes around 12\,s to evolve one model. Then, the parameter steps of models are narrowed down to fine steps around models with minimum $\sigma_{RMS}$. For the fine steps, there are 3 to 5 grid points for each parameter. For fine steps of mass, 0.005 is chosen because it’s the same step size adopted in WDEC. For the other fine steps, they are all chosen to be much smaller than their crude steps. In Table\,\ref{ParaSpace}, there are 6 global parameters and 6 $X_{O}$ parameters. In order to find the minimum fitting error,  the $X_{O}$ parameters are first fixed, and a grid of global parameters are generated. Then the global parameters are fixed, and a grid of $X_{O}$ parameters are generated. The chosen fitting model will be nearly convergent to an optimal fitting model after repeating these two processes for a dozen times. Finally, the optimal model is obtained when $\sigma_{RMS}$ = 0.21\,s (The derived distance is calculated later as 98\,pc). For the additional analysis, the optimal model is found when $\sigma_{RMS}$ = 0.046\,s and its derived distance is 103\,pc. The optimal model parameters of adopted and additional solution are listed in the fifth and last column of Table\,\ref{ParaSpace}, respectively. Although the optimal solution in additional analysis is achieved with a smaller fitting error of 0.046\,s, the solution with fitting error of 0.21\,s is adopted, due to the latter solution results in only one $l$=2 modes. With more high degree modes in the adopted solution, more modes exist in the solution model as well, smaller fitting errors can be achieved, but it is not a reasonable choice comparing to the one with less high degree modes. In addition, in order to explore models with thicker H and He-envelops, ($-log(M_{\rm He}/M_{\rm *})=2$ and $-log(M_{\rm H}/M_{\rm *})=4$, which we did not include in the beginning due to limited computational power), a model grid containing 314\,928 models is calculated. Independent model fittings are performed, yet the adopted solution listed in the fifth column of Table\,\ref{ParaSpace} is still preferred after comparison. On the other hand, in both solutions shown in Table \,\ref{ParaSpace}, close periods 973.734\,s and 967.113\,s are not both $l$=1, meaning they are not likely to be components of a rotational triplet. The full width at half height of the reciprocal of $n\sigma_{RMS}$ is estimated as the error of each optimal value. It is known that modes with $k\gg$1 probe the most external part of the star. Thus the lack of short period modes in our observation reflected in larger errors in the core (h1-w3), compared to the errors estimated for L19-2 \citep{Chen2022}. For L19-2, same models and approach have been applied, but its observations have detected more short period modes, which yields much smaller errors in the core. According to the derived optimal values, J2345 is a relatively massive and hot DAV with a thin H atmosphere.

\begin{figure*}
    \centering
    \includegraphics[width=0.8\textwidth]{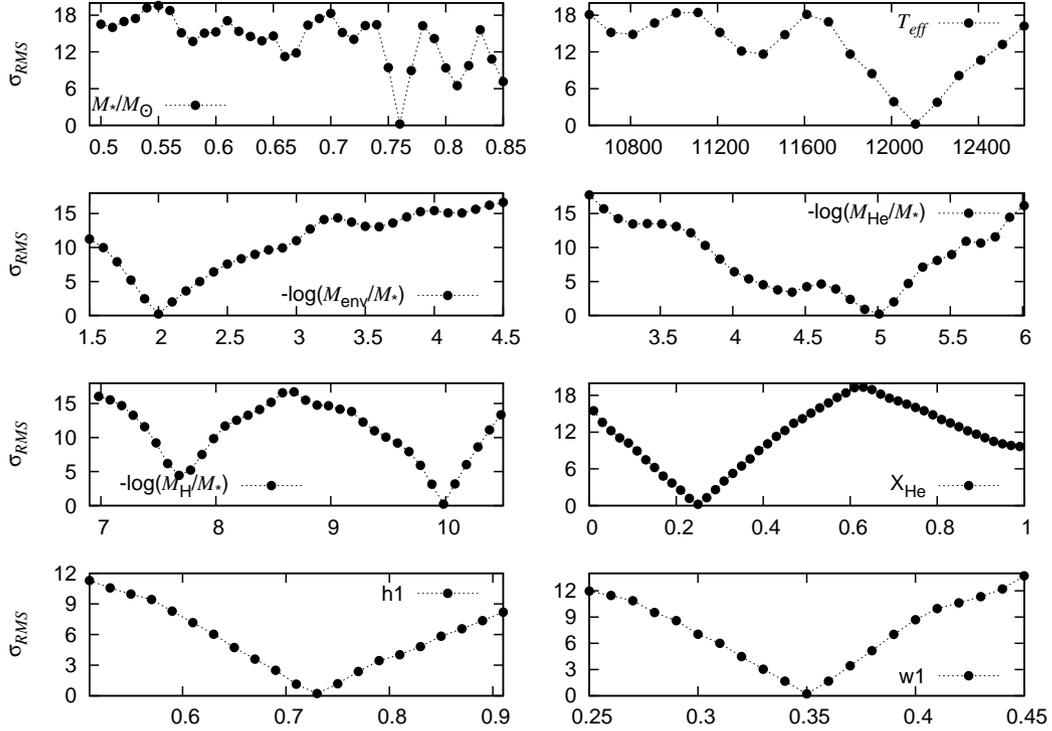}
    \caption{Sensitivity plots of 8 parameters of the optimal model. Subplots are 8 parameter values versus their root mean square residual $\sigma_{RMS}$. }
    \label{fig:dependence}
\end{figure*}

\begin{figure}
    \centering
    \includegraphics[width=0.5\textwidth]{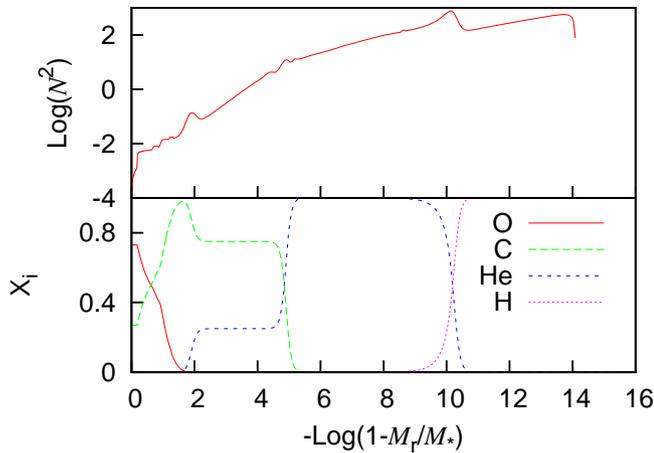}
    \caption{The core composition profiles and corresponding Brunt-V{\"a}is{\"a}l{\"a} frequency for the optimal model. In the lower panel, the abundance of O, C, He and H are plotted in red solid line, green dashed line, blue dashed line and magenta dotted line, respectively. In the upper panel, the vertical axis is the logarithm of the square of buoyancy frequency. Three little bumps are present due to the existence of three element transition gradient zones.}
    \label{fig:2cxln2}
\end{figure}

A colour graph of fitting error $\sigma_{RMS}$ as functions of stellar mass and $T_{\rm eff}$ is displayed in the left panel of Fig.\,\ref{fig:teffmass_envhe}. The colour represents the fitting error $\sigma_{RMS}$ as defined in Eq.\ref{eq:1}. The grid consisting of stellar mass in the range of 0.500\,$M_{\odot}$ and 0.850\,$M_{\odot}$ with a step of 0.005\,$M_{\odot}$ and $T_{\rm eff}$ in the range of 10\,600\,K and 12\,600\,K with step of 10\,K are included in the analysis, which contains 14\,271 DAV models. All of other parameters are fixed to their optimal values in Table \ref{ParaSpace}. The right panel of Fig.\,\ref{fig:teffmass_envhe} shows $\sigma_{RMS}$ as functions of He-layer mass and envelope mass, which involves 7\,676 DAV models. The logarithm of envelope mass ranges from -2.00 to -3.00 with a step of 0.02, and the logarithm of He layer mass ranges from -4.01 to -6.01 with a step of 0.02. 

The sensitivity plots of 8 parameters of the optimal model are shown in Fig.\,\ref{fig:dependence}. Stellar mass, $T_{\rm eff}$, envelope mass, He-layer mass, H-atmosphere mass, He abundance in mixed C/He/H region, h1 and w1 for the O profile versus fitting error $\sigma_{RMS}$ are plotted in different subplots, respectively. Each panel of these parameters is plotted when all other parameters are fixed to the optimal values. The parameter values corresponding to the lowest $\sigma_{RMS}$ indicate the optimal values. In Table\,\ref{ParaSpace}, the mass of the shell of H ranges from $10^{-10}\,M_{\ast}$ to $10^{-5}\,M_{\ast}$. But our fitting is not limited to explore spaces in this range. Depending on the pre-optimal solutions of each iteration in the search for final optimal values of J2345, the initial ranges listed in Table\,\ref{ParaSpace} may be extended if pre-optimal value is near the range limit. For instance, it is shown in Fig.\,\ref{fig:dependence} that H masses less than $10^{-10}\,M_{\ast}$ have been explored as well, since the optimal value is near the limit of initial lower range. Similar case can be found in the mass of the shell of He, spaces as low as $10^{-6}\,M_{\ast}$ are explored as well, exceeding the lower limit of $10^{-5}\,M_{\ast}$.
The core composition profiles and corresponding Brunt-V{\"a}is{\"a}l{\"a} frequency for the optimal model are shown in Fig.\,\ref{fig:2cxln2}, where the bumps are caused by composition gradient in different zones. There are some smaller bumps exist near horizontal axis 1 and 5 as well. After some simple investigations on this matter, we assume that those have something to do with WDEC, rather than anything physical. One should note that this kind of chemical profiles are in contrast with those derived from fully evolutionary computations. The existence of a C-buffer and several features in the chemical profile has been discussed extensively in \citet{Geronimo2019} for a DBV star.

In order to check the self-consistency of the best fit model, the distance derived from asteroseismology analyzed results can be used to do an independent test. For the best-fit model, the luminosity of J2345 is Log(L/L$\odot$) = -2.685. The bolometric magnitude of the optimal model is $M_{bol}$=11.45, using the correction $M_{bol} = M_{bol,\odot} -2.5\times Log(L/L_{\odot})$ and assuming the bolometric magnitude of the Sun $M_{bol,\odot} = 4.74$ \citep{Cox2000}. Since the bolometric correction (BC) in Gaia G band is not provided for J2345 in DR3, the BC in V band is adopted. The BC(V) for DA star model with $T_{\rm eff}$= 12\,000\,K and 13\,000\,K are -0.611 mag and -0.828 mag, respectively, as listed in Table\,1 of \cite{Bergeron1995a}. Therefore, for J2345 with $T_{\rm eff}$= 12\,110\,K, the linear interpolated BC(V) is -0.635 mag. The absolute V magnitude can be derived as $M_{V}=M_{bol}-BC(V)=11.45-(-0.635)=12.085$. By using the equation $d=10^{(m_{V}-M_{V})/5+1}$, the distance corresponding to the optimal model is 93\,pc, given $m_{V}=16.93$ \citep{Lasker2008}. Comparing to the inverse parallax distance of 98\,pc from Gaia DR3, there is $\sim $5\% error in the optimal model derived distance. Furthermore, the results corresponding to the optimal model are in agreement with the parameters obtained from spectral model fitting in section 2.2. These two facts suggest that our asteroseismological results are self consistent.

\begin{figure}
    \centering
    \includegraphics[width=0.55\textwidth]{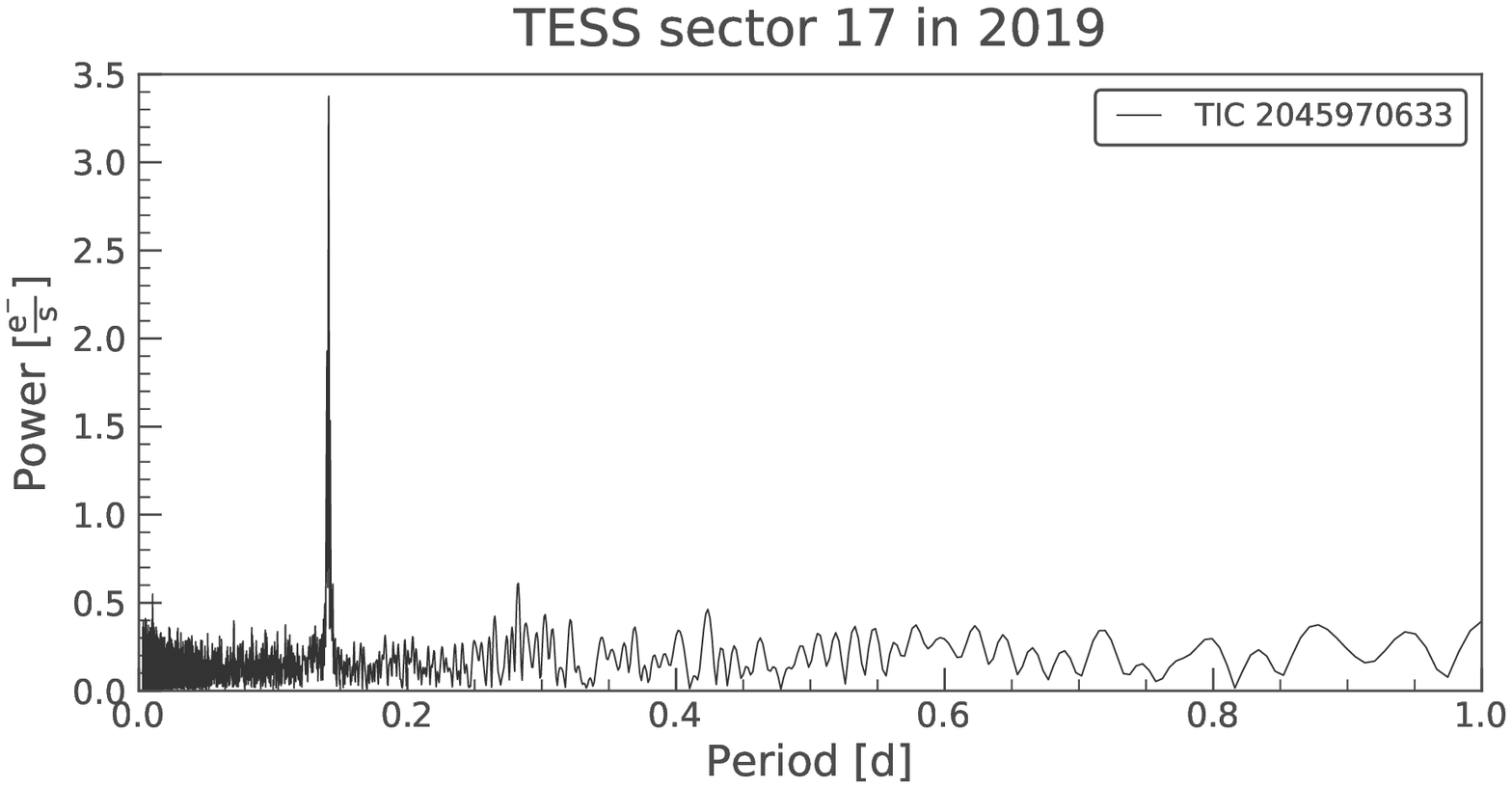}
    \includegraphics[width=0.55\textwidth]{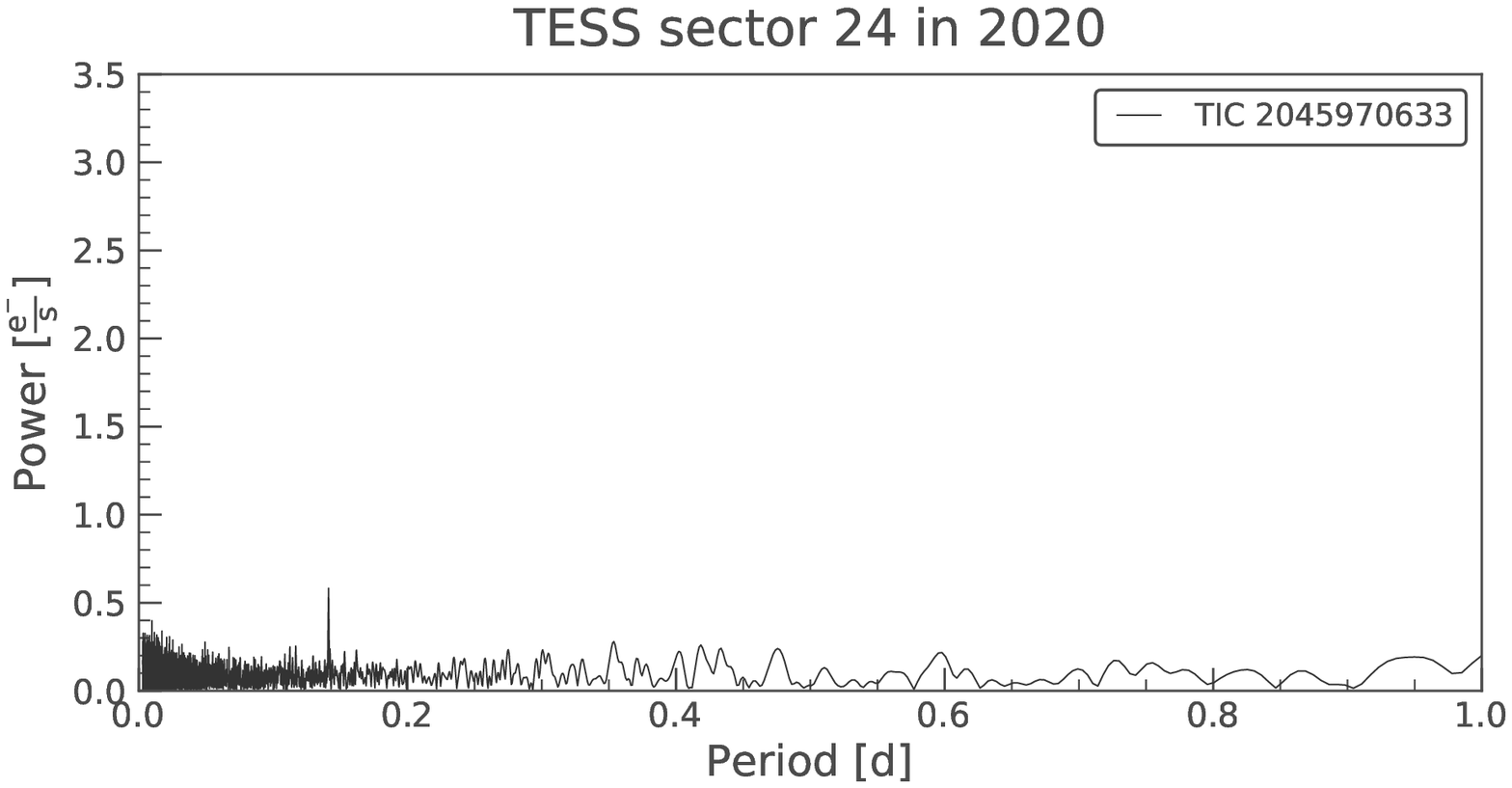}
    \caption{The Lomb-Scargle periodogram of TESS 120\,s observations taken in 2019 and 2020, respectively. The 3.39\,h period becomes much weaker in 2020 than in 2019.}
    \label{fig:tess_LS}
\end{figure}

\section{Rotation period of nearby M dwarf from TESS}
J2345 was observed by TESS in Sector 17 and 24. They are both recorded in 120\,s and 1800\,s exposures. Based on Lomb-Scargle \citep[LS, ][]{Scargle1982} method, a noticeable period $\sim$ 3.39\,h is present in the power spectra of both TESS observations (Figure \ref{fig:tess_LS}), which is not detected in TMTS and SNOVA observations. \cite{Rebassa2019} listed J2345 and another star named Gaia DR2 1999127510436274048 as comoving stars (see Table 1 in their paper), as they are located within about 3$^{\prime\prime}$ and have similar parallaxes and proper motions from Gaia DR2. According to their SED fitting, the $T_{\rm eff}$ of these two stars are 11\,750\,K and 2\,800\,K, respectively, indicating that the comoving star is an M star. Since the image resolution of TESS is $\sim 21^{\prime\prime}$ per pixel and the brightness of M star is comparable to J2345 in TESS red band, the detected 3.39\,h period may originate from the M star. Firstly, the 3.39\,h period is not plausible to be the pulsation period of J2345, as it is too long. Secondly, the power of 3.39\,h period is much weaker in 2020 than in 2019 (see Fig.\,\ref{fig:tess_LS}). Combining the fact that this period is not detected in 2021 observations from TMTS and SNOVA, it is most probable that this period is due to the rotation of M star. The appearance and disappearance of a large dark spot can cause the variation observed in 2019 and 2020. The non-detection in 2021 further rules out the possibility that this period is due to eclipsing of a hidden companion, since the significance of orbital period signal should be stable. To further validate our hypothesis, the folded light curve of TESS observation in 2019 is fed to \texttt{PHOEBE} for modelling. 

\begin{figure}
    \centering
    \includegraphics[width=0.55\textwidth]{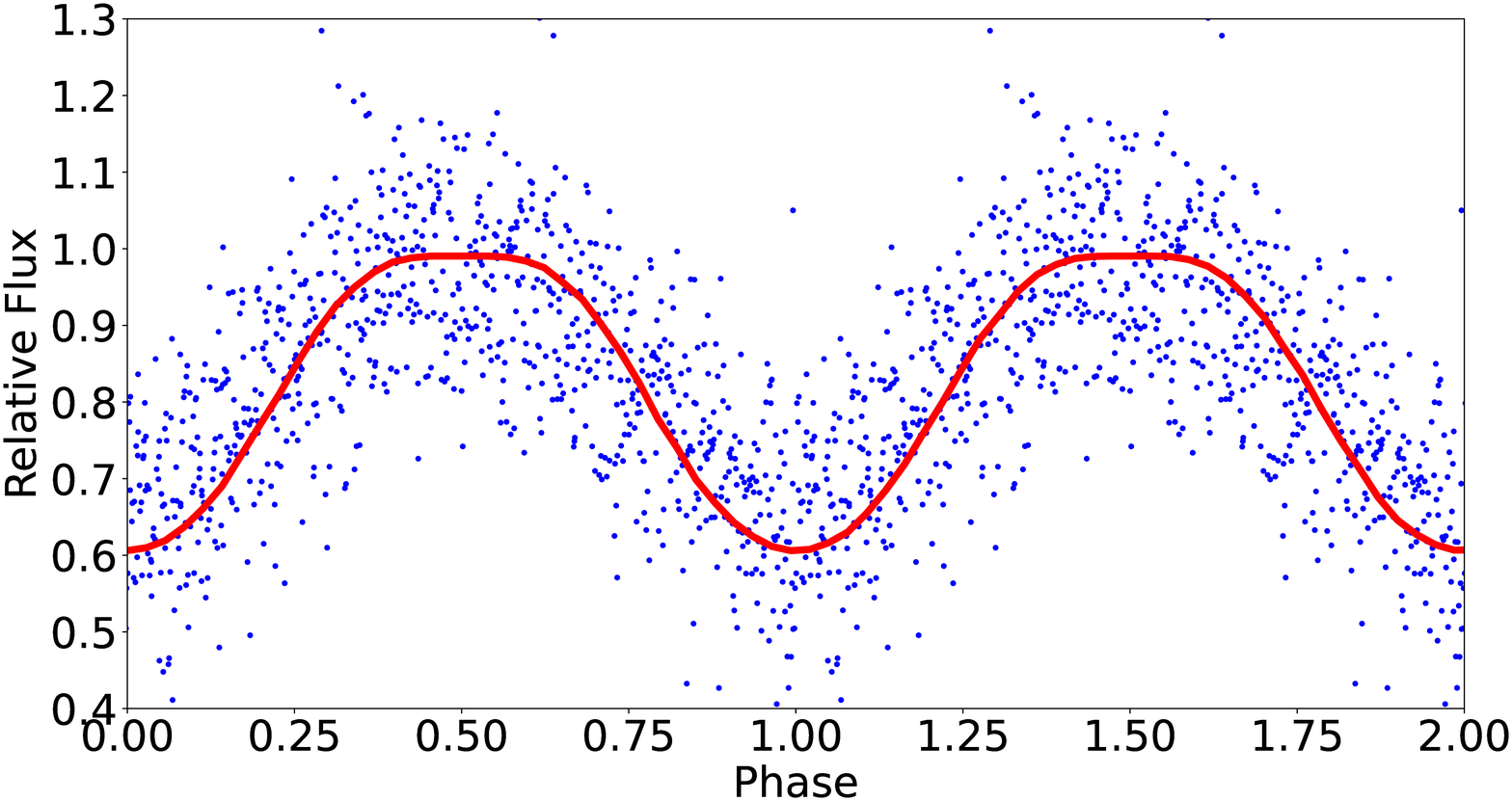}
    \caption{Phase folded TESS light curve modelling with \texttt{PHOEBE}. Blue dots are observed phase-folded TESS light curve in sector 17 with 30\,min exposure, while red line represents the \texttt{PHOEBE} reproduced light curve of a single star with a single large dark spot.}
    \label{fig:spotfitting}
\end{figure}

\begin{figure}
    \centering
    \includegraphics[width=0.6\textwidth]{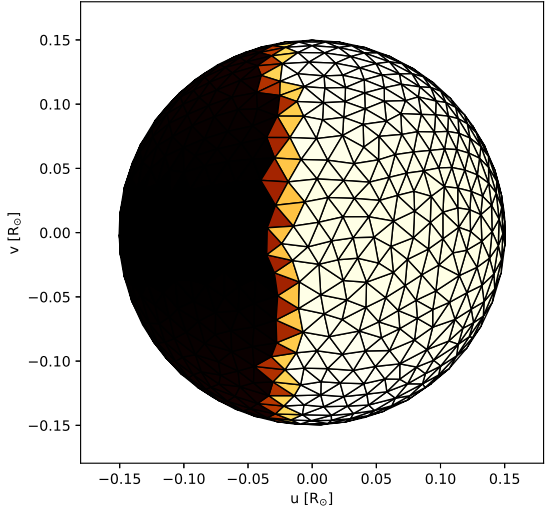}
    \caption{Mesh plot of \texttt{PHOEBE} model of a single star with a single spot, corresponding to the red line light curve at phase=0.75 with inclination angle of 90$^{\circ}$. The star rotates from left to right.}
    \label{fig:spotphases}
\end{figure}

It is known that modelling stellar spots is difficult, as the observed light curve can be well fitted by one large spot or several smaller spots and etc. Currently, \texttt{PHOEBE} does not have spot estimators for single stars. Therefore, we manually tuned the spot parameters to reproduce a similar light curve to mimic the phase-folded light curve of TESS in Sector 17 (Figure\,\ref{fig:spotfitting}). The parameters of the star are set as $T_{\rm eff}$=2\,800\,K, mass=0.21\,M$_{\odot}$, star radius =0.15\,R$_{\odot}$, rotation period=3.39\,h and inclination angle=90$^{\circ}$. The single large spot setting results in: an angular radius of 80$^{\circ}$; the co-latitude of the center of spot with respect to spin axis is 90$^{\circ}$,, longitude of the center of the spot with respect to spin axis is 0$^{\circ}$ and the relative temperature of the spot to the intrinsic temperature is 0.97. The direct image of this spot setting is shown in Fig.\,\ref{fig:spotphases}, corresponding to the model light curve at phase 0.75. It should be noted that this does not necessarily mean this model setting is the real case. It is simply a demonstration to show that the TESS light curve in 2019 can be modelled by the rotation of a large dark spot on the M dwarf.

\section{Discussion}
According to the coordinates from Gaia DR3, the separation between J2345 and the M star is roughly 2.67$^{\prime\prime}$, corresponding to a projected separation of $\sim $15\,AU at 98 parsec away. Based on the parallaxes from Gaia DR3, the distances of the two stars in the line of sight are 94.6288 and 97.7039\,parsec, respectively. Therefore, the actual separation should be at least 3\,parsec, indicating these two stars may be the result of a disrupted wide binary \citep{Oh2017}. There is no third star with similar proper motion at similar distance within 5$^{'}$ in Gaia DR3, thus they do not belong to a large structure. The PanSTARRS images\footnote{\url{http://ps1images.stsci.edu/cgi-bin/ps1cutouts}}
in $z$ and $y$ bands, which have the earliest and latest available images taken 1\,039\,d and 743\,d apart, respectively, are checked as well. However, no obvious relative movements can be observed. 

\begin{figure}
    \centering
    \includegraphics[width=0.5\textwidth]{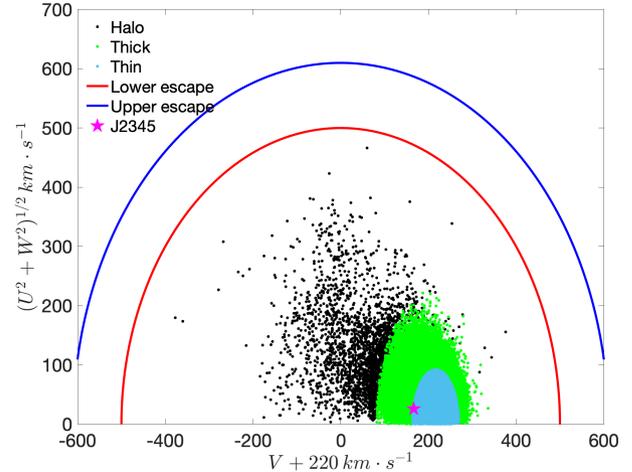}
    \caption{Gaia DR3 selected 1\,kpc sample and J2345 plotted in the V+220\,km$\cdot s^{-1}$ vs. $(U^{2}+W^{2})^{1/2}$\,km$\cdot s^{-1}$ plane. Halo, thick disk, thin disk candidates are plotted in black, green and light blue, respectively. Red and blue lines are lower and upper limits of the escape velocity of the Milky Way. Magenta star marks the location of J2345.}
    \label{fig:Toomre}
\end{figure}

Given the newly released Gaia DR3, it is worth to check whether J2345 is disk or halo WD, since halo WD is rare object of interest. The accurate measurements of J2345 include coordinates of ra=356.2808\,deg, dec=58.2209\,deg, parallax of 10.2350\,mas, proper motion of pmra=28.9359\,mas$\cdot yr^{-1}$, pmdec=5.8859\,mas$\cdot yr^{-1}$. With the radial velocity of -67.7$\pm$17.2\,km$\cdot s^{-1}$ derived from follow-up spectrum with \texttt{WDTOOLS}, the 3-D heliocentric velocity $U$, $V$ and $W$ of J2345 can be calculated following equations listed in \cite{Guo2016}. The probability calculated from $U$=-23.85\,km$\cdot s^{-1}$, $V$=-53.87\,km$\cdot s^{-1}$ and $W$=9.88\,km$\cdot s^{-1}$ suggests that J2345 is likely a thick disk star, not a halo WD (see Fig.\ref{fig:Toomre}). This result is consistent with the disrupted wide binary scenario, as wide binaries in the disk may be disrupted by passing stars, Galactic tides and molecular clouds \citep{Weinberg1987,Jiang2010}. But halo wide binaries are less likely to be disrupted by passing stars and molecular clouds since majority of their lifetime is spent outside the disk. Although some halo wide binaries will travel through the disk, the short duration due to their high spatial velocities makes it less probable to be disrupted \citep{Weinberg1987}.

Since the asterosesimological analysis has provided the accurate mass of J2345, the rough evolution history of this WD can be learnt via stellar evolution models. Firstly, the initial mass at zero age main-sequence (ZAMS) can be calculated as 3.08\,M$_{\odot}$, based on Equation\,\ref{eq:3}, which is derived from Eq.\,5 in \cite{Cummings2018}.
\begin{equation}
M_{i}=(M_{f}-0.184M_{\odot})\div 0.187,  \\
(0.72M_{\odot} < M_{f} < 0.86M_{\odot})
\label{eq:3}
\end{equation}
Where $M_{i}$ is the initial mass, while $M_{f}$ is the final mass. This equation is a MIST-based semi-empirical initial-final mass relation. In Table\,1 of \cite{Renedo2010}, for a WD star with Z=0.01 and M$_{WD}$=0.767 M$_{\odot}$, the lifetime of its main-sequence phase lasts for 0.232\,Gyr. Thus, the main-sequence plus cooling age of J2345 with M$_{WD}=0.760\pm0.005\,M_{\odot}$ is T$_{MS}$+T$_{cooling}\approx$200+700=900\,Myrs. This is less than one per\,cent of the lifetime of the companion M star with $T_{\rm eff}$=2\,800\,K, assuming they formed simultaneously.

On another note, if the initial mass of J2345 at ZAMS is 3.08\,M$_{\odot}$, it should be a B9 or A0 type star \citep{Cox2000}. But the companion of its progenitor is a late M type star. Therefore, it is also possible that there was an accretion process, as binaries formed in the same region tend to have similar spectral types, and the WD mass in accreting system (i.e. mean M$_{WD}\sim$ 0.8\,M$_{\odot}$ for CV) is usually higher than the average mass ($\sim$ 0.6\,M$_{\odot}$) of single WD \citep{Zorotovic2020}.

\section{Summary}
In this work, a newly discovered ZZ ceti J2345 is reported and analyzed. The discovery is the result of searching for variable WDs from the minute-cadence photometric survey TMTS. By cross-matching Gaia EDR3 WD catalog with the TMTS catalog, there are more than a thousand common WD candidates. Each is planned to be visually inspected for variation that can be attributed to WD pulsation, binary eclipsing, and planetesimal transiting. Many interesting sources have been identified, and this is the very first paper to report our findings in TMTS. 

Based on two and five night observations of TMTS and SNOVA, there are 6 pulsating modes identified after prewhitening. Those modes are used to perform the asteroseismological analysis. By adopting the latest version of \texttt{WDEC}, there are 7\,558\,272 WD models evolved for the parameter spaces listed in Table \ref{ParaSpace}. To find the optimal model, parameter steps and spaces are slowly narrowed down repeatedly, in accordance of the fitting result. At last, the optimal model is found and its parameter values are listed in Table\,\ref{ParaSpace}. To test our mode identification, additional analysis of freeing the values of $l$ for all periods has been checked and compared. Based on the comparison and analysis in Section 3.3, we believe our identification is a more reasonable choice. From the asteroseismological result, it can be learnt that J2345 is a massive and relatively hot pulsating WD with moderately thin H atmosphere. The final results are quite consistent with the parameters derived from spectral fitting of follow-up spectrum. In addition, the distance deduced from asteroseismological results is in agreement with the inverse parallax derived distance from Gaia DR3, implying a good self-consistency. 

An evident period of $\sim$3.39\,h is found to exist in two observations of TESS. Given the fact that the significance of this signal weakens considerably between two TESS observations, we propose that this period is attributed to the rotation of dark spot on a comoving M star. The TESS light curve taken in 2019 has been modelled by \texttt{Phoebe} with a single star plus one dark spot model. The result suggests that an synthetic light curve representing a very large dark spot, nearly covered half of the M dwarf's surface can be reproduced to mimic the observed TESS light curve. This implies that the identified period and light curve evolution from TESS data can be explained by stellar spot. But the resulted parameters of the spot should be treated with caution, as the real case may be caused by several spots with different sizes and locations. With the newly released Gaia DR3, the 3-D heliocentric velocity $UVW$ of J2345 is calculated, and the result indicates that J2345 is likely a thick disk star instead of a halo WD. Furthermore, since the initial progenitor mass of WD can be estimated from current mass, its whole evolution history can be learnt. By comparing to a well-studied WD with similar mass, we estimate the lifetime of the main-sequence phase of J2345 being as about 200\,Myrs and current age as 900\,Myrs plus cooling time.

\section*{Acknowledgments}
The authors acknowledge the National Natural Science Foundation of China (NSFC) under grants 12203006, 12033003, 12288102, U1938113, 11773035, 12090044, 11833002 and 11633002. This work is also supported by the Ma Huateng Foundation, the Scholar Program of Beijing Academy of Science and Technology (DZ:BS202002), and the Tencent Xplorer Prize, and CSST projects: CMS-CSST-2021-B03, the Innovation Project of Beijing Academy of Science and Technology (11000023T000002062763-23CB059 and 11000022T000000443055).
 
This work has made use of data from the European Space Agency (ESA) mission {\it Gaia} (\url{https://www.cosmos.esa.int/gaia}), processed by the {\it Gaia} Data Processing and Analysis Consortium (DPAC,
\url{https://www.cosmos.esa.int/web/gaia/dpac/consortium}). Funding for the DPAC has been provided by national institutions, in particular the institutions participating in the {\it Gaia} Multilateral Agreement.
This research has made use of the SIMBAD database, operated at CDS, Strasbourg, France \citep{Wenger2000}.
\section*{DATA AVAILABILITY}
The TESS data used in this article can be accessed via \url{https://archive.stsci.edu/missions-and-data/tess}. The TMTS, SNOVA photometric and 3\,m spectral data can be obtained by contacting the corresponding authors.


\end{document}